# Phase Behaviors of Ionic Liquids Attributed to the Dual Ionic and Organic Nature

Chenyu Tang（唐晨宇）[1,2] and Yanting Wang（王延颋）[1,2*]

[1]CAS Key Laboratory of Theoretical Physics, Institute of Theoretical Physics, Chinese Academy of Sciences, 55 East Zhongguancun Road, P. O. Box 2735, Beijing 100190, China

[2]School of Physical Sciences, University of Chinese Academy of Sciences, 19A Yuquan Road, Beijing 100049, China

**Abstract:** Ionic liquids (ILs), also known as room-temperature molten salts, are composed of pure ions with melting points usually below 100 °C. Because of their low volatility and vast amounts of species, ILs can serve as "green solvents" and "designer solvents" to meet the requirements of various applications by fine tuning their molecular structures. A good understanding of the phase behaviors of ILs is certainly fundamentally important in terms of their wide applications. This review intends to summarize the major conclusions so far drawn on phase behaviors of ILs by computational, theoretical, and experimental studies, illustrating the intrinsic relationship between their dual ionic and organic nature and the crystalline phases, nanoscale segregation liquid phase, ionic liquid crystal phases, as well as phase behaviors of their mixture with small organic molecules.

## 1. Introduction

Ionic liquids (ILs) are a type of salts with low melting points, often below 100 °C, meaning that they tend to remain in the liquid phase at room temperature and are believed to exhibit some unique features because of the strong electrostatic interactions among ions. Typical aprotic ILs are normally composed of small anions and bulky cations with a long alkyl side-chain and a charged head group, as shown in Fig.1, which demonstrates the chemical structure of 1-butyl-3-methylimidazolium chlorine, a typical imidazolium-based IL. Possessing both ionic and organic features, they are believed to have advantageous properties of both organic liquids and inorganic salts, such as good solvation ability and tunability, low melting temperature, good conductivity, wide electrochemical window, thermal and electrochemical stabilities, non-volatility, and non-flammability [1-4]. They are thus regarded as "green" and "engineer" solvents that can be utilized under many industrial circumstances [5-11].

Understanding fundamental properties of ILs, particularly their phase behaviors, is apparently essential to their applications. To investigate their phase behaviors, many computational and experimental methods have been employed to investigate phase behaviors of ILs. Molecular dynamics (MD) simulation has become an important means of studying the structure and dynamics of ILs [12-15] where different modelling methods employing various software packages including GROMACS, NAMD, LAMMPS, etc. [16-18] have been developed. All-atom force fields are commonly used in addressing IL related problems by means of MD simulation [19-21], and the applicability of some commonly used all-atom force fields, including the Amber force field [22], OPLS force field [23], CHARMM force field [24], etc., to IL systems has been verified and the models have been constantly improved to be better applied to ILs [15,25-35]. By considering the polarizable effect at the atomic level, a polarizable model has also been developed to better quantify microscopic structures and dynamics of ILs. [36-38]. Another modelling strategy is to apply a coarse-grained (CG) model in MD simulation, which reduces the cost of computation and renders researchers with longer simulated times than all-atom models. One of the CG methods that are used in the context of ILs is the Multiscale Corse-Graining (MS-CG) method [39-41], which matches the instantaneous forces applied to atoms in the all-atom MD simulation to determine the optimal empirical CG forces between CG sites (atomic groups). By contrast, the Effective Force Coarse-Graining (EF-CG) method directly calculates the effective averaged force between each pair of CG sites (atomic groups) to gain better transferability [42,43]. Other MD methods applicable to study ILs include *ab initio* MD [26,44-46], MD with a polarized force field [47-54], and MD with other CG models [55-60]. Apart from MD simulations, Monte Carlo (MC) simulations [61-63], electronic correlation method [64-67], and Density Functional Theory (DFT) calculations [68-72] are also applied to studying ILs, coming up with many favorable results.

As typical complex liquids, ILs usually have abundant and distinctive phase behaviors beyond the description of simple liquid theories. Some unique phases, including the nanoscale segregation liquid (NSL) phase [73-77], ionic liquid crystal (ILC) phases [78-83], the "partially arrested" glassy phase [84], and the metastable crystal phase [85], were revealed by MD simulation and verified by experiment. Much attention has been aroused since the detailed knowledge of these phases can shed light on the understanding of not only ILs but also other amphiphilic complex liquids. It also provides guidance for industrial utilization of ILs as novel solvents. In this review, we majorly focus on summarizing the phase behaviors of aprotic ILs with alkyl cationic side chains whose number of atomic groups are even. We will show that these behaviors tend to be affected highly by temperature, length of the cationic alkyl side-chains, charge distribution, and nature of cations and anions, indicating that these unique phase behaviors of ILs result from their dual ionic and organic nature [86]. By reviewing these phase behaviors and the associated affecting factors and mechanisms, we aim to inspire further investigations in this direction and invite more researchers to delve into this field to develop a thorough understanding of ILs in the future.

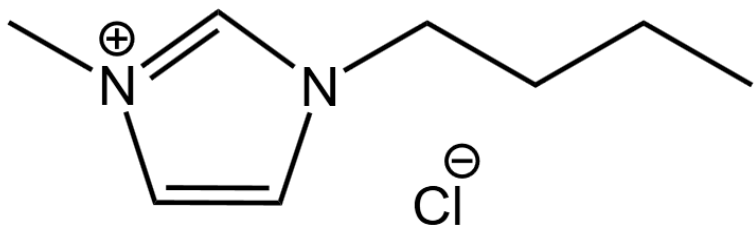

**Figure 1.** Chemical Structure of 1-butyl-3-methylimidazolium chloride.

2. **Dual Ionic and Organic Nature of ILs**

One of the most significant features of ILs that broadens their industrial usage is that ILs inherit both the ionic nature of inorganic salts and the organic nature of organic solvents. They have the advantages of non-flammability, non-volatility, good stability, and conductivity compared to organic solvents, and low melting temperature compared to traditional inorganic salts. By varying cations and anions, millions of available ILs can be produced with various physical and chemical properties [7], which is highly beneficial to meet certain requirements for designated applications. However, since selecting out an appropriate combination of cation and anion via experiment is usually tedious or even unfeasible, it is critical to have a good knowledge of the dual ionic and organic nature of ILs to help achieving computer-aided systematic design of ILs. It has been well acknowledged that the competition between electrostatic and van der Waals (VDW) interactions is the key factor characterizing the ionic and organic nature [87,88], so analyzing these interactions in ILs by means of MD simulations with suitable force fields should be informative.

In a recent work, Shi and Wang [86] performed a series of all-atom MD simulations for four representative ILs (1-butyl-3-methylimidazolium nitrate ([BMIM][$NO_3$]), 1-butyl-3-methylimidazolium tetrafluoroborate ([BMIM][$BF_4$]), 1-butyl-3-methylimidazolium hexafluorophosphate ([BMIM][$PF_6$]), and 1-butyl-3-methylimidazolium bis(trifluoromethylsulfonyl)imide ([BMIM][$Tf_2N$])), and compared them with three molecular systems with different charge distributions: a typical molten inorganic salt (molten sodium chloride, NaCl), a strongly polar liquid (Dimethyl sulfoxide, DMSO), and a weakly polar liquid (toluene). The dual ionic and organic nature of ILs can be depicted from the viewpoint of the cage energy landscape (CEL) by the obtained forces, vibrational force constants, intrinsic electric fields, cohesive energies, and cage energies.

By introducing the concept of ion cage [89-97], which indicates how each ion is surrounded by several counter-ions in the first coordination shell, along with its counterpart of molecular cage [98-102] for molecular liquids, it is possible to depict the structures and dynamics of various types of liquids since the cage volume corresponds to density and the cage stability reflects dynamics. From cage structure, it is then possible to determine CEL by calculating the ensemble-averaged local energy landscape as a function of the dislocation of the central ion from the cage center. The curvature, slope, and depth of CEL can be determined as the force constant $k$, force $F$, and activation energy $E_a$, which is the average energy of a particle required to climb over the energy barrier and escape the cage calculated by using the harmonic approximation (Fig. 2a).

Shi and Wang [86] employed the first moment of the vibrational density of state (VDOS) to qualitatively describe the average characteristic frequency of intermolecular vibrational modes, defined as [103]

$$\langle \omega \rangle = \int_0^{\omega_c} \omega I(\omega) \mathrm{d}\omega / \int_0^{\omega_c} I(\omega)\, \mathrm{d}\omega \qquad (1)$$

where $\omega$ is the frequency and $I(\omega)$ the VDOS. The total, VDW, and electrostatic forces, respectively, are compared in all the investigated liquids, and a liquid-phase cage energy $U_{\mathrm{cage}}$ is defined as the average potential energy between an ion and a counter-ion in its ion cage to characterize the local ion-ion interaction in the liquid.

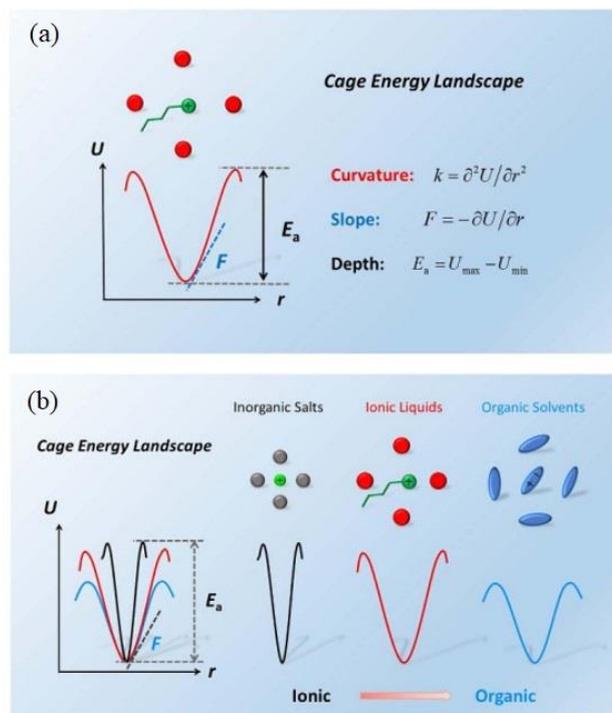

**Figure 2.** (a) Cage energy landscape characterized by curvature, slope, and depth, corresponding to force constant, force, and activation energy experienced by molecules, respectively. (b) Schematic illustration of cage structures and cage energy landscape in inorganic salts, ionic liquids, and organic solvents. The cage energy landscape of inorganic salts is deep and steep, whereas that of ionic liquids is still deep but much more gently. Organic solvents and ionic liquids have a similar slope and curvature near the minimum of the cage energy landscape, but the depths for the organic solvents are much lower. Reprinted with permission from R. Shi and Y. Wang, Sci. Rep. 6, 19644 (2016).

From these approaches, a conclusion has been drawn that similar molecular size, geometry, and component lead to comparable VDW forces in organic ions and organic molecules. They also weaken the electrostatic interactions in ILs because of charge delocalization and charge transfer. The cage energy of ILs induced by electrostatic interactions is drastically different from organic liquids. These findings indicate that the VDW interactions, which dominate the intermolecular forces and vibrational force constants, characterize the organic nature of RTILs, resulting in a similar geometry near the minimum of the CEL to organic liquids; whilst the cage energy, or the depth of the CEL, can characterize their ionic nature (Fig. 2b). Such a mechanism explains the similar and dissimilar characteristics between ILs and organic liquids, and it clarifies the blurry dual ionic and molecular nature of ILs whereas the corresponding microscopic mechanism provides new insights into their phase behaviors.

3. **Nanoscale Segregation Liquid (NSL) Phase**

A unique phase behavior in ionic liquids that can be explained by the dual ionic and organic nature of ILs is the nanoscale segregation liquid (NSL) phase of ILs, which is different from either simple liquid phase or liquid crystal (LC) phase. The structure was first discovered by MD simulation and later verified by experiment [74,76,104,105]. The discovery of the NSL phase may possibly boost new industrial applications of ILs, and the microscopic mechanism of NSL shall aid the engineering of ILs [76]. The significance of this discovery is enhanced by the fact that the NSL exists in most IL systems with an amphiphilic cation, independent of a specific choice of the anion.

In a series of MS-CG MD simulations on [C$_n$MIm][NO$_3$] with $n$ = 4–12, Wang, et al. [74,76,106] observed a tail-aggregation phenomenon and later referred to it as the nanoscale segregation liquid (NSL) phase. It has been found that, in a certain temperature range, while the whole liquid is macroscopically homogeneous, the tail groups of the cationic alkyl side chains form nanoscale nonpolar tail domains microscopically segregated with the continuous polar network formed by charged anions and cationic head groups. This phenomenon was later confirmed experimentally [104,107,108] by using optical heterodyne-detected Raman-induced Kerr effect spectroscopy (OHDRIKES), Neutron Diffraction, X-ray diffraction, electrospray ionization mass spectrometry (ESI-MS), and NMR measurements, all providing solid evidence for the existence of the NSL phase.

To quantify the degree of tail aggregation in the NSL phase, the Gaussian-like Heterogeneity Order Parameter (HOP) is defined as [76]

$$h = \frac{1}{N} \sum_{i,j=1}^{N} \exp\left(-\frac{r_{ij}^2}{2\sigma^2}\right) \quad (2)$$

to quantify the spatial heterogeneity for identical sites, where $r_{ij}$ is the distance between sites $i$ and $j$, and $\sigma = L/N^{1/3}$ with $L$ being the side length of the cubic simulation box and $N$ being the total number of identical sites.

Both the radial distribution functions (RDFs) and HOPs of the simulated systems show that, compared with ILs with short alkyl side chains, which is basically in the simple liquid phase, the tail groups distribute quite heterogeneously in ILs with an intermediate alkyl side chain length. The charged domain, constituted by anions and cationic head groups, is dominated by the electrostatic interactions, whereas the neutral cationic tail groups form a nonpolar tail domain with the VDW interactions among tail groups. It has also been discovered that such aggregation behavior is temperature-sensitive, which can be understood through analyzing the tail domain diffusion in ILs [76]. Investigating the behavior of tail domains with increasing temperature has also revealed that the transition of ILs from NSL to simple liquid is characterized by the collective behavior of tail groups. Nevertheless, such a collective behavior is passively induced by the steric repulsion from the continuous polar network, as concluded by a further study on ILs with an external electric field applied that the repulsion from the polar part is the major cause of the aggregation of the nonpolar tail groups [77].

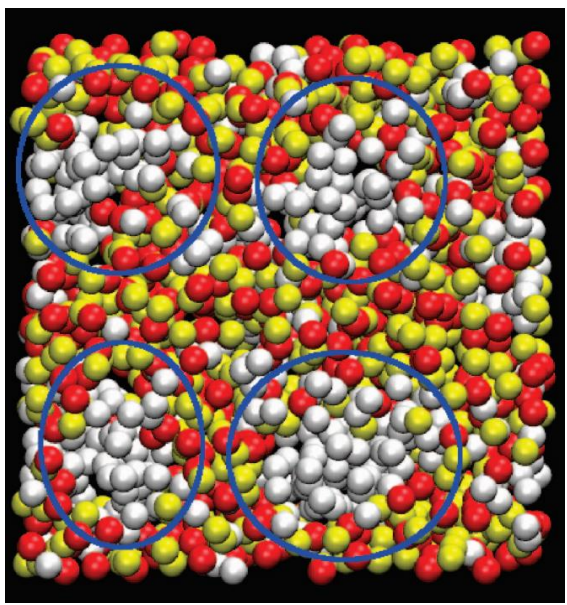

**Figure 3.** Snapshot illustrating the NSL phase. The white spheres represent the cationic terminal groups, the gold spheres represent the cationic head groups, and the red spheres represent the anions. The ellipses in blue indicate the approximate positions of the nonpolar tail domains. Reprinted with permission from Y. Wang, W. Jiang, T. Yan, and G. A. Voth, Acc. Chem. Res. 40, 1193 (2007). Copyright 2007 American Chemical Society.

### 4. Ionic Liquid Crystal (ILC) Phases

While ILs with an intermediate cationic alkyl side-chain length can form the unique NSL phase, it has been revealed by experiment that ILs with longer alkyl chains can easily form ILC phases, most of which are smectic [80,81,83,109-114]. It is therefore critical to unveil the mechanism for ILs to transform from NSL to ILC phases when the length of alkyl side chains increases. To study this phase transition, Ji et al. [82] and Li et al. [115] adopted the EF-CG model for ILs [43] to perform MD simulations on $[C_n MIm][NO_3]$ with $n$ = 6–22 [82] and 12–24 [115], respectively. Both simulations indicate that the phase transition from NSL to ILC exists when the side-chain length is increased.

Ji et al. [82] has simulated 512 ion pairs with an isotropic barostat and cubic simulation box. The HOP values reveal that the IL systems go through a structural transition when the side-chain length increases beyond 14, and the HOP of tail groups decreases drastically from $C_{14}$ to $C_{16}$ (Fig. 4a). It not only coincides with previous experimental studies [83] but also is verified through RDFs of the investigated systems. The orientation correlation function (OCF) for side chains has also been introduced, which is the ensemble-averaged correlation between the orientations of two side chains as a function of the distance between the center-of-masses (COMs) of cations:

$$C(r) = \left\langle \left[ 3\left(\hat{u}(\vec{r}_i) \cdot \hat{u}(\vec{r}_j)\right)^2 - 1 \right] \cdot \delta(\vec{r} - \vec{r}_i + \vec{r}_j)/2 \right\rangle \quad (3)$$

where $\hat{u}(\vec{r}_i)$ is the unit vector pointing from the head to the tail of cation $i$.

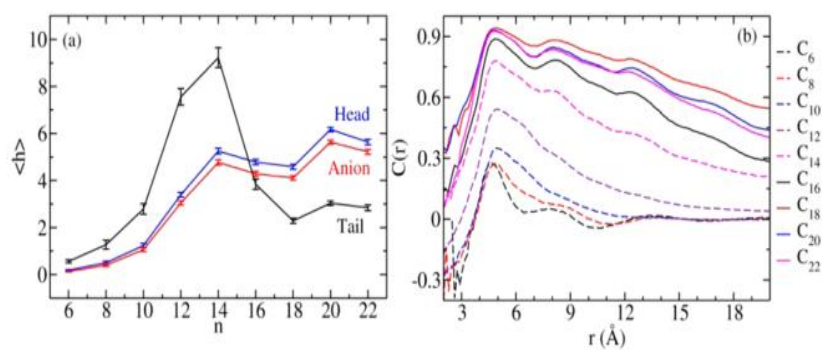

**Figure 4.** (a) HOP for cationic head groups, anions, and cationic tail groups. (b) OCF for side chains. Reprinted with permission from Y. Ji, R. Shi, Y. Wang, and G. Saielli, J. Phys. Chem. B 117, 1104 (2013). Copyright 2013 American Chemical Society.

As shown in Fig. 4b, the maximum value of OCF increases with the side-chain length from 6 to 14, and stays close to 1 from 16 to 22, indicating that the side chains of neighboring cations change their relative spatial feature from aggregated to parallel-aligned. It can also be observed that, with increasing distance, the OCF values for $C_6$ to $C_{14}$ gradually decrease to zero and those for $C_{16}$ to $C_{22}$ gradually decrease to a finite value. This provides further evidence that the transformation is a phase transition from the NSL phase to the ILC-like phase since, in the NSL phase, side chains do not have long-range correlations, whereas they have a long-range order in the ILC-like phase. It has been suggested that the increment of the VDW interactions might be the major cause for such a transition.

Li et al.'s work [115] further investigated this phase transition and interpreted it from the perspective of percolation phase transition. Larger systems (4096 ion pairs) have been simulated with an anisotropic barostat allowing the simulation box size in three dimensions to change independently and an isotension-isothermal ensemble allowing the simulation box to change its shape. In the configurations equilibrated by simulating annealing procedures, two side chains are considered "connected" when their COM distance is less than 0.72 nm and at the same time the twist angle between them is less than 30°. A set of "connected" side chains is then defined as a cluster. In IL systems with an intermediate side-chain length, such as $C_{12}$, the side chains have weak tendencies to form clusters (Fig. 5a). With increasing side-chain length, in the $C_{16}$ system, more and larger clusters formed locally because of the stronger tendency of parallel alignment of side chains, but the orientations of the clusters are still random (Fig. 5b). For $C_{22}$, side chains are in parallel globally and the largest cluster is comparable to the size of the simulated box, and the second-largest cluster is comparatively very small (Fig. 5c), implying the occurrence of a percolation phase transition [116,117]. By analysing the size difference between the largest and second-largest clusters (Fig. 5d), it has been determined that the phase transition happens at $C_{18}$ where both the average cluster size and the correlation length reach their maxima of around 500 and 0.42, respectively, with large fluctuations (Fig. 5e and Fig. 5f).

In the above simulations, although a CG model has been adopted, the simulated IL systems still suffer from limited equilibration time and finite-size effect. Therefore, the above results render future studies with even larger temporal and spatial simulation scales to refine the microscopic mechanism of the phase transition in ILs from the NSL state to the ILC state.

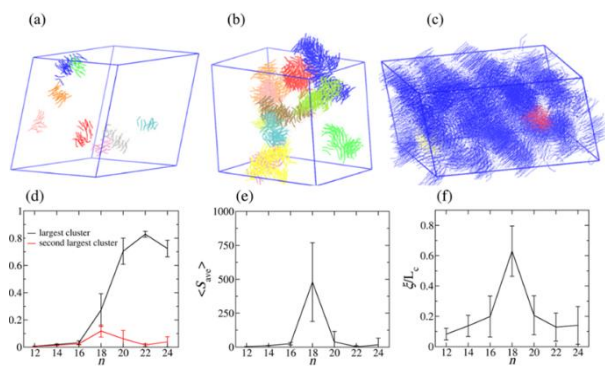

**Figure 5.** Percolation phase transition. (a–c) Several largest clusters in $C_{12}$, $C_{16}$ and $C_{22}$ systems, respectively, with the largest cluster colored blue. (a) The largest cluster in $C_{12}$ is very small and not well aligned. (b) The largest cluster in $C_{16}$ is larger and better aligned in parallel, but still local with little orientation correlation between clusters. (c) The largest cluster in $C_{22}$ almost fills in the whole simulation box, indicating that the majority of the side chains are globally aligned in parallel and well connected. (d) Normalized sizes of the largest and second-largest clusters for all systems. (e) Average cluster size versus side-chain length. After the percolation phase transition, the largest cluster is not counted in the calculation of the average cluster size. (f) Correlation length versus side-chain length. For a finite system, both the average cluster size and the correlation length reach their maxima at the phase transition point. The correlation length is directly related to the average cluster size. Reprinted with permission from S. Li and Y. Wang, Sci. Rep. 9, 13169 (2019).

## 5. Solid-Solid Phase Transition and Melting Transition

Apart from varying side-chain lengths of the ILs, it has also been observed in computational and experimental studies that phase transitions can exist when varying temperatures or artificially tuning partial charges of ions [118,119]. By MD simulations with the EF-CG model conducted on [$C_{16}$MIm][$NO_3$], Saielli et al. [118] have discovered that, with increasing temperature, the phase of IL can transit firstly from crystalline to smectic A (SmA) ILC, and then from ILC to NSL, which is similar to experimental results conducted on [$C_n$MIm][$BF_4$] [81,120]. The HOP provides direct evidence for the existence of such transitions since the HOP of the tail groups (CG sites E) increases drastically between 470 K and 505 K, corresponding to the phase transition from crystal to ILC, and decreases after 550 K, from ILC to NSL (Fig. 6). The snapshots representing the development from ILC to crystal at 480 K and from NSL to ILC at 550 K, respectively, are illustrated in Fig. 7.

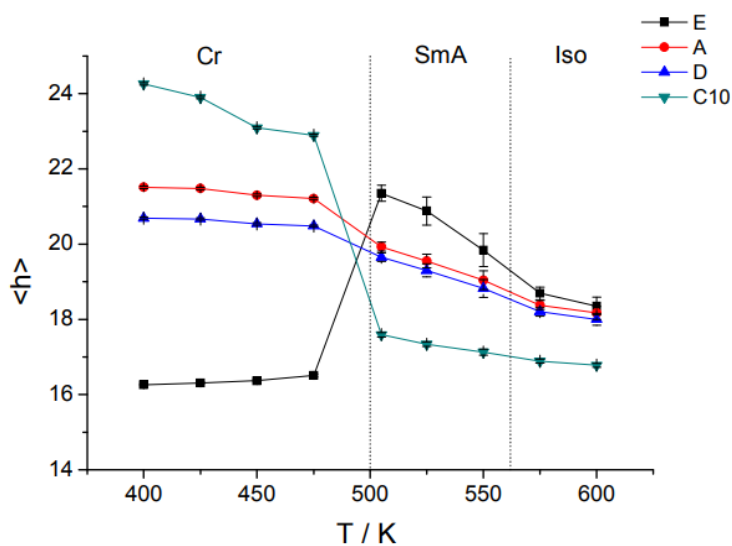

**Figure 6.** HOPs for some CG sites as a function of temperature. The error bars are almost invisible since they are smaller than the size of the markers. Cr represents the crystal phase, SmA the smectic A phase, and Iso the nanoscale segregation liquid phase. Reprinted with permission from G. Saielli, A. Bagno, and Y. Wang, J. Phys. Chem. B 119, 3829 (2015). Copyright 2015 American Chemical Society.

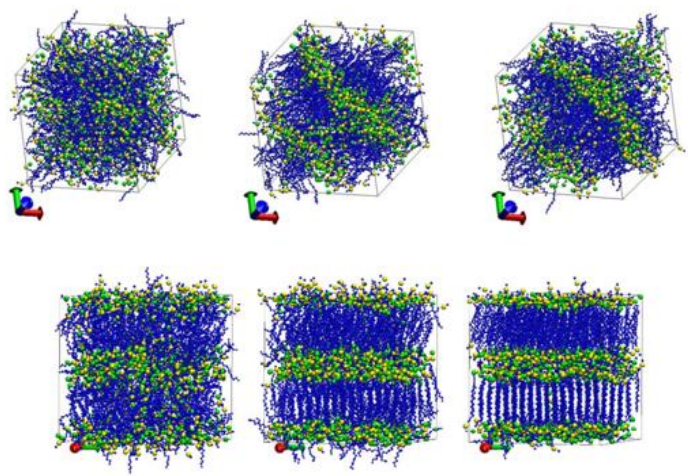

**Figure 7.** (Upper Row) Snapshots of the simulation box during the transition from the NSL phase to the ILC phase taken after 0, 48, and 72 ns when the temperature was set at 505 K. (Lower Row) Snapshots of the simulation box during the transition from the ILC to the crystal phase taken after 0, 5, and 20 ns when the temperature was set at 480 K. Reprinted with permission from G. Saielli, A. Bagno, and Y. Wang, J. Phys. Chem. B 119, 3829 (2015). Copyright 2015 American Chemical Society.

Similar phase behavior can also be observed while tuning partial charges [119]. The phase transitions are observed and determined by calculating the orientational order parameter (OOP) and the translational order parameter (TOP), which are commonly used in describing the LC phase transitions. It is thus possible for the phase diagram to be determined as a function of temperature and partial charges (Fig. 8). This study suggests that the existence of ILC phases is strongly increased by the total charge of the ions. When the partial charges on the CG sites are rescaled by a factor lower than 0.9, the ILC phase is absent and the crystal directly melts into the NSL phase. For the systems with larger partial charges, the thermal range for a stable ILC phase is significantly increased. The HOP of the NSL phase also suggests that the nanoscale segregation is largely affected by the partial charges: the IL systems tend to be homogenous with low partial charges while nanosegregate with larger partial charges. The IL systems melted from the ILC phase have a higher degree of nanosegregation, as measured by the HOP, than that melted directly from crystal at the same temperature. These studies indicate that the increase of either the alkyl chain length or the total charge of the cation head group and anion intensify the competition between hydrophobic and electrostatic interactions, which enlarges the existence ranges on the phase diagram for both the NSL phase and the ILC phase.

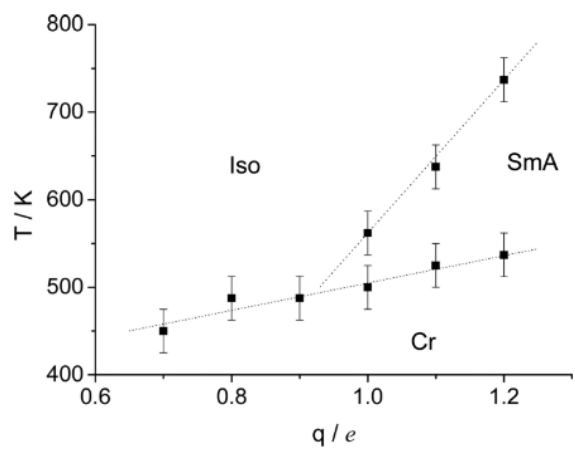

**Figure 8.** Phase diagram of the IL system simulated by the EF-CG model. Cr represents the crystal phase, SmA the smectic A phase, and Iso the nanoscale segregation liquid phase. Reprinted with permission from G. Saielli and Y. Wang, J. Phys. Chem. B 120, 9152 (2016). Copyright 2016 American Chemical Society.

Cao et al. [85] have investigated the phase behavior of $[C_n\text{MIm}][NO_3]$ ($n = 4-12$) during heating by manually constructing the initial crystal structures that constitute parallel polar layers composed of cationic head groups and anions as well as nonpolar regions composed of cationic alkyl side chains in between. The initial configuration is then heated by all-atom MD simulation with the AMBER force field.

During the heating process, a solid-solid phase transition has been found below the melting point, manifested by the kinks on the caloric curves shown in Fig. 8. The jumps of the potential energies at the solid-solid phase transition points are much smaller than those at melting transition points. The $C_4$ system exhibits distinctiveness compared with others due to its weaker VDW interaction coming from the shorter side chains. The difference between such phase transitions of $C_4$ and $C_8$ are shown in Fig. 9. The TOPs and OOPs indicate that the conformations of alkyl chains of the $C_4$ system are not as ordered as those with longer side chains.

Further increasing the simulation temperature indicates that the melting phase transition of the investigated systems consists of two steps except $C_6$, which has only one step. A metastable state exhibits during the melting transition when the crystalline solid phase is transformed into the NSL phase (for $C_4$, $C_6$, and $C_8$) or SmA ILC phase (for $C_{10}$ and $C_{12}$). The snapshots of the metastable states of $C_4$ and $C_8$ systems are shown in Fig. 8b. The metastable state tends to be more stable along with a higher melting temperature when the alkyl chain becomes longer. These features of melting transitions can be comprehended by the competition between the free energy contributions of the effective VDW interaction in the nonpolar regions (or EF1 for abbreviation) and that of the effective electrostatic interaction in the polar layers (or EF2 for abbreviation). The existence of the unique solid-solid phase behavior of ILs can thus be attributed to the dual ionic and organic nature of ILs.

As the side-chain length increases, EF1 becomes stronger and thus the system possesses ordered conformations during solid-solid phase transitions. During the melting process, the imbalance between EF1 and EF2 at the melting point leads to an uneven melting process of polar layers and nonpolar regions, which causes the metastable phase to appear. For the $C_4$ system, EF1 is weaker than EF2 at the melting point of 315 K, so the alkyl side chains lose the crystalline order before the polar layers do. For systems containing $C_8$ to $C_{12}$, longer alkyl side chains lead to stronger EF1, and thus the increase of the melting temperatures, at which the alkyl side chains keep their orientations uniformly for a certain time whilst polar layers drift from their original lattice positions. For the $C_6$ system, the free energy contributions from the two interactions are well balanced, resulting in the absence of the metastable phase.

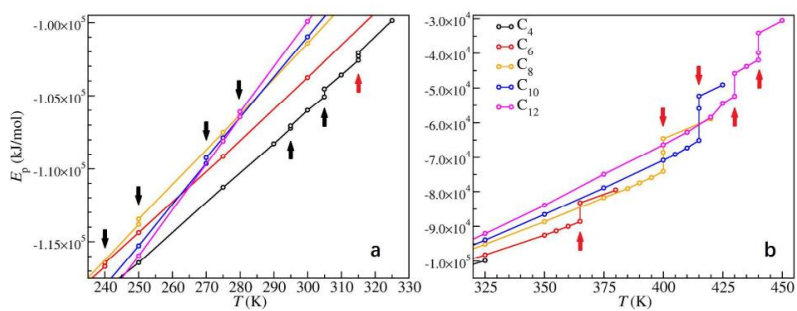

**Figure 9.** Caloric curves during heating. (a) From $T$ = 240 K to 325 K. (b) From $T$ = 325 K to 450 K. The solid-solid phase transition points are marked by black arrows, and the melting transition points are marked by red arrows. Reprinted with permission from W. Cao, Y. Wang, and G. Saielli, J. Phys. Chem. B 122, 229 (2018). Copyright 2018 American Chemical Society.

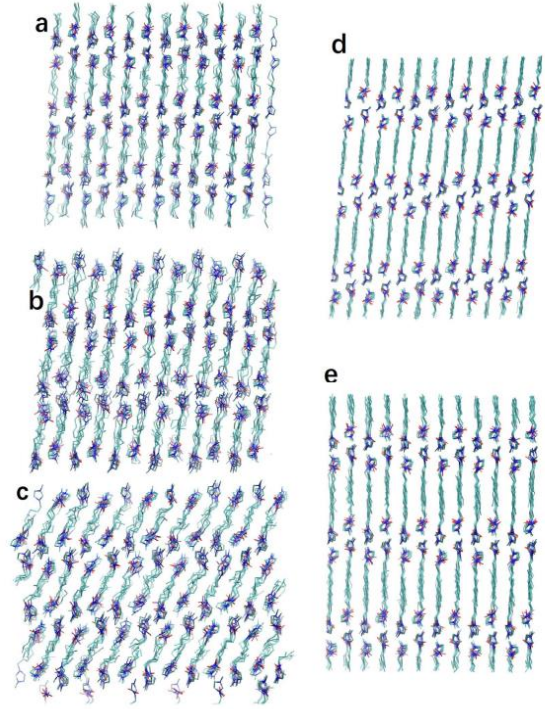

**Figure 10.** Snapshots of the structures before and after the solid-solid phase transitions. (a) $C_4$ at 295 K before the phase transition. (b) $C_4$ at 300 K after the first but before the second phase transition. (c) $C_4$ at 305 K after the phase transition. (d) $C_8$ at 250 K before the phase transition. (e) $C_8$ at 250 K after the phase transition. All these snapshots are taken from the [100] direction. Reprinted with permission from W. Cao, Y. Wang, and G. Saielli, J. Phys. Chem. B 122, 229 (2018). Copyright 2018 American Chemical Society.

**6. Partially Arrested Glassy State**

The Self-Consistent Generalized Langevin Equation (SCGLE) theory has been used to investigate the dynamics of molten salts, and such implementation has predicted the existence of the so-called dynamically arrested state when the larger ions are still in the fluid state while the smaller counter-ions are arrested in the glassy state [121]. The SCGLE theory puts forward a simple equation for the asymptotic value of the mean-squared displacement of species $\alpha$, which is defined as $\gamma_\alpha \equiv \lim_{t \to \infty} \left\langle \left( \Delta R^{(\alpha)} \right)^2 \right\rangle$, and the equation reads as [122]:

$$\frac{1}{\gamma_\alpha} = \frac{1}{3(2\pi)^3} \int d^3k k^2 \left\{ \lambda \left[ \lambda + k^2 \gamma \right]^{-1} \right\}_{\alpha\alpha} \times \left\{ c\sqrt{n} S \lambda \left[ S\lambda + k^2 \gamma \right]^{-1} \sqrt{n} h \right\}_{\alpha\alpha} \qquad (4)$$

where $S$ is the matrix of partial structure factors, $h$ and $c$ are the Ornstein-Zernike matrices of total and direct correlation functions, respectively. The matrix $\sqrt{n}$ is defined as $[\sqrt{n}]_{\alpha\beta} \equiv \delta_{\alpha\beta}\sqrt{n_\alpha}$, and $\lambda(k)$ as $\lambda_{\alpha\beta}(k) = \delta_{\alpha\beta}\left[ 1 + \left(k/k_c^{(\alpha)}\right)^2 \right]^{-1}$ where $k_c^{(\alpha)} = 8.17/\sigma_\alpha$ with $\sigma_\alpha$ being the diameter of the investigated particle. The theoretical prediction for molten salts indicates that there are partially arrested states in the dynamic arrest line (F-G region in Fig. 11a) where only one species of ions are arrested in the glassy state while the counter-ions are still in the liquid state.

Ramírez-Gonzalez et al. [84] have applied the SCGLE theory to investigating the [C$_2$MIm][BF$_4$] IL and found that the IL system is possibly in the partially arrested glassy state. Anions' value $\gamma_{anion}^{-1}$ varies from zero at higher temperatures than cations' value $\gamma_{cation}^{-1}$, which suggests that the IL is in the F-G region with the anions arrested while the cations being liquid. As a consequence, although at short times, the mobility of BF$_4^-$ is higher than that of C$_2$MIm$^+$ due to the free flight in the ballistic regime, it is observed that the diffusion of the smaller BF$_4^-$ becomes slower than the larger C$_2$MIm$^+$ in the diffusive regime, as observed in the all-atom MD simulations with the Amber force field [22] (Fig. 11b). By analyzing the RDFs between cations and anions, Ramírez-Gonzalez and co-workers have discovered that the anions behave as a typical Wigner glass whose mean distance of the constituents is very large.

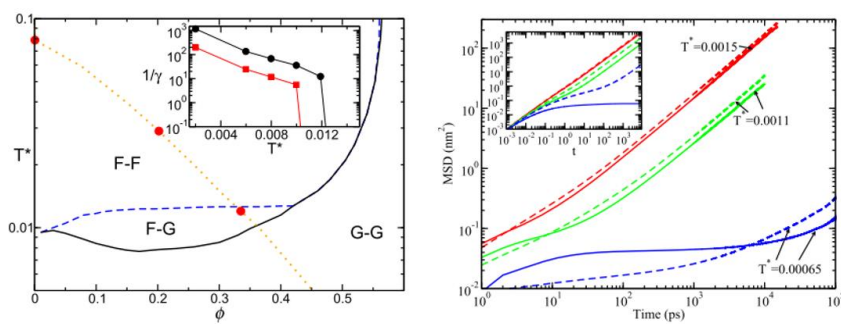

**Figure 11.** (Left) Arrest lines predicted by the SCGLE theory for the primitive model with a size asymmetry adequate for the simulated IL (1:3.5). The fluid region is labelled as F-F. The G-G region corresponds to fully arrested states. The partially arrested phenomenon occurs in the F-G region. The inset shows the values of the parameter 1/γα following the isobaric trajectory with a pressure of 1 atm indicated by the dotted line. Circles correspond to anions (smaller particles) and squares to cations (larger particles). (Right) MSDs for [EMIM][BF$_4$]. Dashed lines represent cations and solid lines represent anions. The inset shows the theoretical calculation results of the MSDs over the isobaric trajectory (big circles in the left panel). The first two temperatures are located in the F-F region and the last one inside the F-G region. Solid and dashed lines correspond to anions and cations, respectively. Reprinted from P. E. Ramirez-Gonzalez, L. E. Sanchez-Diaz, M. Medina-Noyola, and Y. Wang, J. Chem. Phys. 145, 191101 (2016) with the permission from AIP Publishing.

## 7. Liquid-liquid Phase Separation of IL/Benzene Mixture

Aside from pure IL systems, the phase behaviors of IL mixtures have also attracted much attention. Specifically, since ILs are found to exhibit dual ionic and organic nature and amphiphilic features, the mixtures of ILs and organic molecules have been widely investigated [123-131]. Unlike ILs directly solved in many small organic solvents, such as acetonitrile, dichloromethane, and chloroform [132], ILs mixed with benzene exhibit liquid-liquid phase separation. Holbrey et al. [126] observed experimentally that aromatic molecules, such as benzene and toluene, are soluble in imidazolium-based ILs and that liquid-liquid phase separations occur in the mixtures of ILs and aromatic liquids. Later studies on the solubility of aromatic molecules in different ILs also indicate that electrostatic interactions between cations and anions can be of much decisiveness in understanding the phase separation [123,124,128-131].

To understand the microscopic mechanisms of the above liquid-liquid phase separation, Li et al. [133] performed both NMR experiments and all-atom MD simulations on a set of mixtures of benzene and viologen bistriflimide salts [C$_m$bpC$_n$)$_2$][CF$_3$SO$_2$)$_2$N] with $m$ and $n$ being the numbers of carbon atoms on the two alkyl side chains. The viologen salts are crystal at room temperature but also exhibit ILC and NSL phases at higher temperatures, as common ILs do [133].

Both NMR experiments and MD simulations confirm that, when mixed with benzene, the salt absorbs benzene molecules to form a sponge-like phase with benzene inserted in the nonpolar domains of the NSL-like structure compose of nonpolar alkyl chains surrounded by the continuous polar network formed by anions and charged cationic head groups (Fig. 12b). This mixture phase coexists with liquid benzene or crystalline viologen-based IL whichever is excessive. The upper boundaries separating the coexisting phases depend linearly on the cationic alkyl chain length of the IL (Fig. 12a) because larger volume of nonpolar domains can absorb more benzene molecules. The lower boundaries also exhibit linear dependency on the side-chain length corresponding to the minimum amount of benzene required to liquidize the salt, which increases proportionally with the cationic side-chain length.

Although many other investigations [123,124,126,129,131] have suggested that π–π or ion-π interactions may contribute to such kind of liquid-liquid phase separations, the MD simulations have indicated that they do not exist in the above benzene/viologen-based IL mixture systems, and the benzene molecules reside inside the nonpolar domains only due to their planar and nonpolar molecular feature. Further studies along this direction may investigate the influence of the alkyl chain length, the cation and anion types, and the temperature on the liquid-liquid phase separation behaviors of the mixtures composed of various ILs and small organic molecules and develop a general theoretical framework for this kind of phase behaviors.

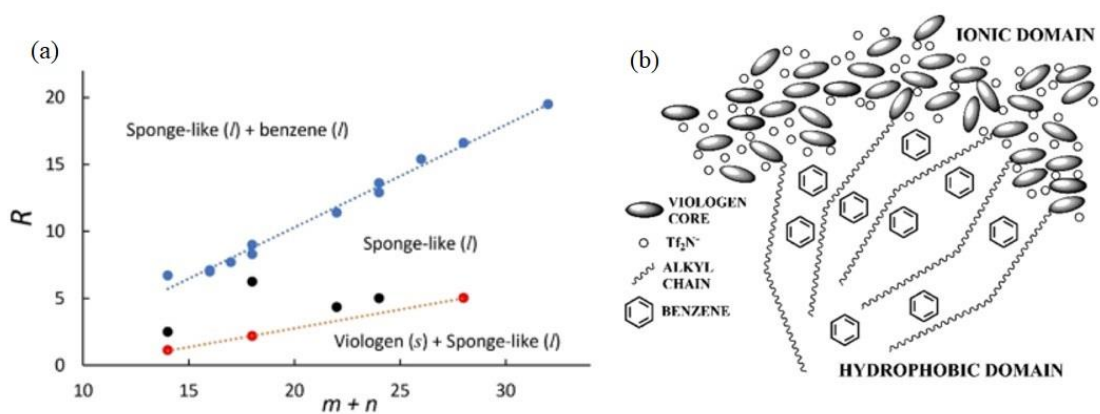

**Figure 12.** (a) Phase diagram of the mole ratio $R = N_{benzene}/N_{viologen}$ vs $m + n$, the total number of carbon groups in the cationic alkyl chains, measured by experiment. The blue symbols represent the coexistence line between liquid benzene and sponge-like liquid phase, the red symbols represent the coexistence line between solid viologen salt and sponge-like liquid phase, and the black symbols represent intermediate state points with only the sponge-like liquid phase present. (b) Schematic representation of the sponge-like phase. The typical nanoscale segregation normally observed in ionic liquids is reproduced in the sponge-like phase after benzene molecules are absorbed in the hydrophobic regions. For the sake of clarity, only a few alkyl chains are schematically drawn. Reprinted with permission from S. Li, N. Safari, G. Saielli, and Y. Wang, J. Phys. Chem. B 124, 7929 (2020). Copyright 2020 American Chemical Society.

8. **Other Phase Behaviors of ILs**

ILs mixed with water constitute an important topic of investigation. Many MD simulations on the mixtures of water with imidazolium-based ILs have focused on studying their structural characteristics, analyzing the RDFs and spatial distribution functions (SDFs), whereas many others have focused majorly on depicting the mixture by analyzing hydrogen bonds or investigating the effect of alkyl chain length on their dynamics [134-144]. Some investigators have also addressed the significance of unique structural features. Jiang et al. [145] studied the influence of water molecules to the NSL structure of [$C_8$MIm][$NO_3$], and water molecules were found to be inserted inside the polar network of the IL. Abe and coworkers [146-149] further determined the existence of aggregations of water molecules in IL systems through discovering the percolation limit that Bernardes et al. [150] predicted through MD simulations. They also brought up a hypothesis that the so-called "water pockets" exist in the mixture. Bystrov et al. [134] have confirmed the existence of nanoscale structures in IL/water mixtures. By using MD simulation and the experimental Pulsed Field Gradient-Stimulated Echo (PFGSTE) method, they have discovered that other than forming "water pockets" or "nanodroplets", water molecules tend to distribute within the hydrophilic region in the IL system. All of these studies indicate that the NSL structure of ILs plays an important role in the mixture of IL and water.

Another set of mixing systems undergoing intense investigations is the binary IL mixtures. It has been widely accepted that, by varying cationic head groups, their side-chain length, and the nature of anions, the NSL structures of ILs can be fine-tuned [104,151-154]. There is, however, an alternative option of mixing two or more types of ILs with different chemical structures [155]. Cosby et al. [156] have thoroughly studied how tuning the composition of binary IL mixtures can affect the mesoscale organization and dynamics of the mixing system with the example of [$C_8$MIm][$BF_4$] and [$C_2$MIm][$BF_4$] mixture. Through detailed X-ray scattering, neutron scattering, and MD simulation studies, Cosby have discovered that increasing the mole ratio of [$C_2$MIm][$BF_4$] can lead to the increase of static dielectric permittivity, $\varepsilon_s$, which is the consequence of a transition in the mesoscale morphologies of the binary mixing system.

Unique phase separation behaviors are observed in many other mixing systems as well. In a computational study of a ternary electrolyte mixture of tetramethylene glycol dimethyl ether (tetraglymeor $G_4$), [$C_2$MIm][$BF_4$], and Li salts (LiNO$_3$ and LiI), Fuladi et al. [157] have discovered that resembling the phase separation of IL mixtures, the ternary mixture tends to separate into two domains with one majorly consisted of ionic species and the other consisted of only $G_4$ molecules. This phase separation has been analyzed by changing volume fraction of $G_4$ and ILs, salt concentration, and temperature. It is believed that the phase separation process is driven primarily by entropy and thus temperature sensitive, which slows the diffusive dynamics of Li$^+$ ions. This study will inspire future investigation and engineering of IL electrolytes that may improve the performance of Li batteries.

9. Conclusions

ILs are room-temperature organic salts that possess both ionic and organic characteristics, which leads to some distinctive phase behaviors both in the context of pure ILs and its mixture with other liquids. Previous theoretical, computational, and experimental studies on phase behaviors of ILs have suggested the existence of many phases in this amphiphilic liquid, whose phase behaviors can be fine-tuned by manipulating thermodynamic conditions and molecular structures, such as charge distribution, length of cationic alkyl side chains, and temperature. Mixing with other molecules can vary their phase behaviors as well.

With an intermediate alkyl cationic side-chain length, ILs exhibit a unique NSL state which is macroscopically isotropic but microscopically phase separated into a polar network and nonpolar domains. The ILC phase appears when the alkyl side chains are sufficiently long. For the crystalline solid of ILs, when increasing temperature, the solid-solid phase transition may occur and the melting transition from crystal to IL or ILC sometimes encounters a metastable state. At a certain thermal condition, an IL system may stay in the partially arrested glassy state. Liquid-liquid phase separation can occur when ILs mixed with some sorts of small organic molecules.

By increasing alkyl side-chain length, ILs go over a percolation phase transition from the NSL phase to the ILC phase due to the increasing VDW interactions among alkyl side chains to transform the conformation of side-chain bundles from nanoscale aggregation to globally parallel alignment. Most of the phase behaviors of aprotic ILs can be qualitatively understood by the competition between the effective free energy contributions from the collective VDW interactions among alkyl side chains and the electrostatic interactions among charged anions and cationic head groups.

It can be noted that the dual ionic and organic nature of ILs plays an important role in determining the unique phase behaviors of ILs. The studies mentioned in this review thus pave the way for future theoretical works investigating phase behaviors of ILs and their mixtures from this perspective. Through analyzing the free energy contributions from electrostatic and VDW interactions, analytical models may be viable to provide a unified mechanism to explain the existing computational and experimental results and provide researchers and engineers a proper guidance to finely tune the physical and chemical properties of ILs in favor of certain industrial applications.

**Corresponding Author**

*E-mail: wangyt@itp.ac.cn. Phone: +86 10-62648749.

**Acknowledgement**

This work was supported by the National Natural Science Foundation of China (Nos. 11774357, 22011530390, 12047503) and the Chinese Academy of Sciences (No. QYZDJ-SSW-SYS01).

**References**

[1] R. D. Rogers and K. R. Seddon, Science **302**, 792 (2003).
[2] M. Gaune-Escard and G. M. Haarberg, *Molten salts chemistry and technology* (John Wiley & Sons, 2014).
[3] J. S. Wilkes and M. J. Zaworotko, Journal of the Chemical Society, Chemical Communications, 965 (1992).
[4] J. S. Wilkes, Green Chemistry **4**, 73 (2002).
[5] M. Armand, F. Endres, D. R. MacFarlane, H. Ohno, and B. Scrosati, Materials for sustainable energy: a collection of peer-reviewed research and review articles from Nature Publishing Group, 129 (2011).
[6] V. I. Pârvulescu and C. Hardacre, Chemical Reviews **107**, 2615 (2007).
[7] N. V. Plechkova and K. R. Seddon, Chemical Society Reviews **37**, 123 (2008).

[8] A. E. Somers, P. C. Howlett, D. R. MacFarlane, and M. Forsyth, Lubricants **1**, 3 (2013).
[9] T. Torimoto, T. Tsuda, K. i. Okazaki, and S. Kuwabata, Advanced Materials **22**, 1196 (2010).
[10] F. Van Rantwijk and R. A. Sheldon, Chemical reviews **107**, 2757 (2007).
[11] T. Welton, Chemical reviews **99**, 2071 (1999).
[12] B. Bhargava, S. Balasubramanian, and M. L. Klein, Chemical communications, 3339 (2008).
[13] P. Hunt, Molecular simulation **32**, 1 (2006).
[14] R. M. Lynden-Bell, M. G. Del Popolo, T. G. Youngs, J. Kohanoff, C. G. Hanke, J. B. Harper, and C. C. Pinilla, Accounts of chemical research **40**, 1138 (2007).
[15] M. S. Kelkar and E. J. Maginn, The Journal of Physical Chemistry B **111**, 4867 (2007).
[16] S. Plimpton, Journal of computational physics **117**, 1 (1995).
[17] H. J. Berendsen, D. van der Spoel, and R. van Drunen, Computer physics communications **91**, 43 (1995).
[18] J. C. Phillips *et al.*, Journal of computational chemistry **26**, 1781 (2005).
[19] J. N. Canongia Lopes, J. Deschamps, and A. A. Pádua, The journal of physical chemistry B **108**, 2038 (2004).
[20] Y. Wang, H. Pan, H. Li, and C. Wang, The Journal of Physical Chemistry B **111**, 10461 (2007).
[21] B. Bhargava and S. Balasubramanian, The Journal of chemical physics **127**, 114510 (2007).
[22] J. Wang, R. M. Wolf, J. W. Caldwell, P. A. Kollman, and D. A. Case, Journal of computational chemistry **25**, 1157 (2004).
[23] W. L. Jorgensen, D. S. Maxwell, and J. Tirado-Rives, Journal of the American Chemical Society **118**, 11225 (1996).
[24] K. Vanommeslaeghe *et al.*, Journal of computational chemistry **31**, 671 (2010).
[25] A. Bagno, F. D'Amico, and G. Saielli, Journal of molecular liquids **131**, 17 (2007).
[26] B. Bhargava and S. Balasubramanian, Chemical physics letters **417**, 486 (2006).
[27] B. Bhargava and S. Balasubramanian, The Journal of chemical physics **123**, 144505 (2005).
[28] B. L. Bhargava, M. L. Klein, and S. Balasubramanian, ChemPhysChem **9**, 67 (2008).
[29] J. N. Canongia Lopes, A. A. Pádua, and K. Shimizu, The Journal of Physical Chemistry B **112**, 5039 (2008).
[30] N. A. Denesyuk and J. D. Weeks, The Journal of chemical physics **128**, 124109 (2008).
[31] R. Lynden-Bell and T. Youngs, Molecular Simulation **32**, 1025 (2006).
[32] N. M. Micaelo, A. M. Baptista, and C. M. Soares, The Journal of Physical Chemistry B **110**, 14444 (2006).
[33] C. Schröder and O. Steinhauser, The Journal of chemical physics **128**, 224503 (2008).
[34] T. G. Youngs, M. G. Del Pópolo, and J. Kohanoff, The Journal of Physical Chemistry B **110**, 5697 (2006).
[35] T. G. Youngs and C. Hardacre, ChemPhysChem **9**, 1548 (2008).
[36] T. Yan, C. J. Burnham, M. G. Del Pópolo, and G. A. Voth, The Journal of Physical Chemistry B **108**, 11877 (2004).
[37] T. Yan, Y. Wang, and C. Knox, The Journal of Physical Chemistry B **114**, 6905 (2010).
[38] T. Yan, Y. Wang, and C. Knox, The Journal of Physical Chemistry B **114**, 6886 (2010).
[39] S. Izvekov, A. Violi, and G. A. Voth, The Journal of Physical Chemistry B **109**, 17019 (2005).
[40] S. Izvekov and G. A. Voth, The Journal of chemical physics **123**, 134105 (2005).
[41] S. Izvekov and G. A. Voth, The Journal of Physical Chemistry B **109**, 2469 (2005).
[42] Y. Wang, S. Feng, and G. A. Voth, Journal of chemical theory and computation **5**, 1091 (2009).

[43]     Y. Wang, W. G. Noid, P. Liu, and G. A. Voth, Physical Chemistry Chemical Physics **11**, 2002 (2009).
[44]     M. G. Del Pópolo, J. Kohanoff, and R. M. Lynden-Bell, The Journal of Physical Chemistry B **110**, 8798 (2006).
[45]     M. G. Del Pópolo, R. M. Lynden-Bell, and J. Kohanoff, The Journal of Physical Chemistry B **109**, 5895 (2005).
[46]     M. H. Ghatee and Y. Ansari, The Journal of chemical physics **126**, 154502 (2007).
[47]     J. L. Banks, G. A. Kaminski, R. Zhou, D. T. Mainz, B. Berne, and R. A. Friesner, The Journal of chemical physics **110**, 741 (1999).
[48]     D. Bedrov, J.-P. Piquemal, O. Borodin, A. D. MacKerell Jr, B. Roux, and C. Schröder, Chemical reviews **119**, 7940 (2019).
[49]     R. Chelli, P. Procacci, R. Righini, and S. Califano, The Journal of chemical physics **111**, 8569 (1999).
[50]     J. Chen, D. Hundertmark, and T. J. Martínez, The Journal of chemical physics **129**, 214113 (2008).
[51]     L. R. Olano and S. W. Rick, Journal of computational chemistry **26**, 699 (2005).
[52]     S. Patel and C. L. Brooks III, Molecular Simulation **32**, 231 (2006).
[53]     S. W. Rick and S. Stuart, Reviews in computational chemistry **18**, 89 (2003).
[54]     M. Kohagen, M. Brehm, J. Thar, W. Zhao, F. Muller-Plathe, and B. Kirchner, J Phys Chem B **115**, 693 (2011).
[55]     B. L. Bhargava, R. Devane, M. L. Klein, and S. Balasubramanian, Soft Matter **3**, 1395 (2007).
[56]     H. A. Karimi-Varzaneh, F. Muller-Plathe, S. Balasubramanian, and P. Carbone, Phys Chem Chem Phys **12**, 4714 (2010).
[57]     C. Merlet, M. Salanne, and B. Rotenberg, The Journal of Physical Chemistry C **116**, 7687 (2012).
[58]     A. Moradzadeh, M. H. Motevaselian, S. Y. Mashayak, and N. R. Aluru, Journal of Chemical Theory and Computation **14**, 3252 (2018).
[59]     M. Salanne, Physical Chemistry Chemical Physics **17**, 14270 (2015).
[60]     L. I. Vazquez-Salazar, M. Selle, A. H. De Vries, S. J. Marrink, and P. C. Souza, Green Chemistry **22**, 7376 (2020).
[61]     F. Bresme and J. Alejandre, The Journal of chemical physics **118**, 4134 (2003).
[62]     J. K. Shah, J. F. Brennecke, and E. J. Maginn, Green Chemistry **4**, 112 (2002).
[63]     W. Shi and E. J. Maginn, The Journal of Physical Chemistry B **112**, 2045 (2008).
[64]     E. I. Izgorodina, M. Forsyth, and D. R. MacFarlane, Physical chemistry chemical physics **11**, 2452 (2009).
[65]     E. I. Izgorodina, Physical Chemistry Chemical Physics **13**, 4189 (2011).
[66]     R. S. Santiago, G. R. Santos, and M. Aznar, Fluid Phase Equilibria **278**, 54 (2009).
[67]     S. Tsuzuki, H. Tokuda, K. Hayamizu, and M. Watanabe, The Journal of Physical Chemistry B **109**, 16474 (2005).
[68]     K. Fujii, T. Fujimori, T. Takamuku, R. Kanzaki, Y. Umebayashi, and S.-i. Ishiguro, The Journal of Physical Chemistry B **110**, 8179 (2006).
[69]     B. G. Janesko, Physical Chemistry Chemical Physics **13**, 11393 (2011).
[70]     K. Karu, A. Ruzanov, H. Ers, V. Ivaništšev, I. Lage-Estebanez, and J. M. Garcia de la Vega, Computation **4**, 25 (2016).
[71]     S. A. Katsyuba, E. E. Zvereva, A. Vidiš, and P. J. Dyson, The Journal of Physical Chemistry A **111**, 352 (2007).
[72]     Y. Zhang, H. He, K. Dong, M. Fan, and S. Zhang, Rsc Advances **7**, 12670 (2017).
[73]     C. Chiappe, Monatshefte für Chemie - Chemical Monthly **138**, 1035 (2007).
[74]     Y. Wang, W. Jiang, and G. A. Voth, in *Ionic Liquids IV* (2007), pp. 272.
[75]     Y. Wang and G. A. Voth, J Am Chem Soc **127**, 12192 (2005).

[76]	Y. Wang and G. A. Voth, The Journal of Physical Chemistry B **110**, 18601 (2006).
[77]	H.-Q. Zhao, R. Shi, and Y.-T. Wang, Communications in Theoretical Physics **56**, 499 (2011).
[78]	G. Casella, V. Causin, F. Rastrelli, and G. Saielli, Liquid Crystals **43**, 1161 (2016).
[79]	G. Casella, V. Causin, F. Rastrelli, and G. Saielli, Physical Chemistry Chemical Physics **16**, 5048 (2014).
[80]	C. M. Gordon, J. D. Holbrey, A. R. Kennedy, and K. R. Seddon, Journal of Materials Chemistry **8**, 2627 (1998).
[81]	J. D. Holbrey and K. R. Seddon, Journal of the Chemical Society, Dalton Transactions, 2133 (1999).
[82]	Y. Ji, R. Shi, Y. Wang, and G. Saielli, The Journal of Physical Chemistry B **117**, 1104 (2013).
[83]	C. K. Lee, H. W. Huang, and I. J. Lin, Chemical Communications, 1911 (2000).
[84]	P. E. Ramirez-Gonzalez, L. E. Sanchez-Diaz, M. Medina-Noyola, and Y. Wang, J Chem Phys **145**, 191101 (2016).
[85]	W. Cao, Y. Wang, and G. Saielli, J Phys Chem B **122**, 229 (2018).
[86]	R. Shi and Y. Wang, Sci Rep **6**, 19644 (2016).
[87]	H. Weingärtner, Angewandte Chemie International Edition **47**, 654 (2008).
[88]	D. Chandler, J. D. Weeks, and H. C. Andersen, Science **220**, 787 (1983).
[89]	M. G. Del Pópolo and G. A. Voth, The Journal of Physical Chemistry B **108**, 1744 (2004).
[90]	Z. Hu and C. J. Margulis, Proceedings of the National Academy of Sciences **103**, 831 (2006).
[91]	X. Huang, C. J. Margulis, Y. Li, and B. J. Berne, Journal of the American Chemical Society **127**, 17842 (2005).
[92]	H. A. Karimi-Varzaneh, F. Müller-Plathe, S. Balasubramanian, and P. Carbone, Physical Chemistry Chemical Physics **12**, 4714 (2010).
[93]	M. Kohagen, M. Brehm, J. Thar, W. Zhao, F. Müller-Plathe, and B. Kirchner, The Journal of Physical Chemistry B **115**, 693 (2011).
[94]	T. I. Morrow and E. J. Maginn, The Journal of Physical Chemistry B **106**, 12807 (2002).
[95]	R. Shi and Y. Wang, The Journal of Physical Chemistry B **117**, 5102 (2013).
[96]	S. Zahn, J. Thar, and B. Kirchner, The Journal of chemical physics **132**, 124506 (2010).
[97]	Y. Zhang and E. J. Maginn, The journal of physical chemistry letters **6**, 700 (2015).
[98]	R. Lynden-Bell, D. Hutchinson, and M. Doyle, Molecular Physics **58**, 307 (1986).
[99]	R. Lynden-Bell and W. A. Steele, The Journal of Physical Chemistry **88**, 6514 (1984).
[100]	M. J. Polissar, The Journal of Chemical Physics **6**, 833 (1938).
[101]	E. Rabani, J. D. Gezelter, and B. Berne, The Journal of chemical physics **107**, 6867 (1997).
[102]	D. A. Turton, J. Hunger, A. Stoppa, A. Thoman, M. Candelaresi, G. Hefter, M. Walther, R. Buchner, and K. Wynne, Journal of Molecular Liquids **159**, 2 (2011).
[103]	T. Fujisawa, K. Nishikawa, and H. Shirota, The Journal of chemical physics **131**, 244519 (2009).
[104]	A. Triolo, O. Russina, H.-J. Bleif, and E. Di Cola, The Journal of Physical Chemistry B **111**, 4641 (2007).
[105]	Y. Wang, W. Jiang, T. Yan, and G. A. Voth, Accounts of chemical research **40**, 1193 (2007).
[106]	Y. Wang and G. A. Voth, Journal of the American Chemical Society **127**, 12192 (2005).
[107]	S. Raju and S. Balasubramanian, The Journal of Physical Chemistry B **114**, 6455 (2010).
[108]	D. Xiao, L. G. Hines Jr, S. Li, R. A. Bartsch, E. L. Quitevis, O. Russina, and A. Triolo, The Journal of Physical Chemistry B **113**, 6426 (2009).
[109]	K. V. Axenov and S. Laschat, Materials **4**, 206 (2011).

[110] K. Goossens, K. Lava, P. Nockemann, K. Van Hecke, L. Van Meervelt, K. Driesen, C. Görller‐Walrand, K. Binnemans, and T. Cardinaels, Chemistry – A European Journal **15**, 656 (2009).
[111] E. Guillet, D. Imbert, R. Scopelliti, and J.-C. G. Bünzli, Chemistry of materials **16**, 4063 (2004).
[112] P. H. Kouwer and T. M. Swager, Journal of the American Chemical Society **129**, 14042 (2007).
[113] G. F. Starkulla, S. Klenk, M. Butschies, S. Tussetschläger, and S. Laschat, Journal of Materials Chemistry **22**, 21987 (2012).
[114] V. Causin and G. Saielli, Journal of Materials Chemistry **19**, 9153 (2009).
[115] S. Li and Y. Wang, Sci Rep **9**, 13169 (2019).
[116] M. Sahimi, (London, 1994).
[117] D. Stauffer and A. Aharony, (1992).
[118] G. Saielli, A. Bagno, and Y. Wang, J Phys Chem B **119**, 3829 (2015).
[119] G. Saielli and Y. Wang, J Phys Chem B **120**, 9152 (2016).
[120] C. J. Bowlas, D. W. Bruce, and K. R. Seddon, Chemical communications, 1625 (1996).
[121] L. E. Sanchez-Diaz, A. Vizcarra-Rendon, and R. Juarez-Maldonado, Phys Rev Lett **103**, 035701 (2009).
[122] R. Juárez-Maldonado and M. Medina-Noyola, Physical Review E **77**, 051503 (2008).
[123] M. Blesic, J. N. C. Lopes, A. A. Padua, K. Shimizu, M. F. C. Gomes, and L. P. N. Rebelo, The Journal of Physical Chemistry B **113**, 7631 (2009).
[124] M. Deetlefs, C. Hardacre, M. Nieuwenhuyzen, O. Sheppard, and A. K. Soper, The Journal of Physical Chemistry B **109**, 1593 (2005).
[125] G. Gonfa, M. A. Bustam, N. Muhammad, and S. Ullah, Journal of Molecular Liquids **238**, 208 (2017).
[126] J. D. Holbrey, W. M. Reichert, M. Nieuwenhuyzen, O. Sheppard, C. Hardacre, and R. D. Rogers, Chemical Communications, 476 (2003).
[127] J. Łachwa, I. Bento, M. T. Duarte, J. N. C. Lopes, and L. P. Rebelo, Chemical communications, 2445 (2006).
[128] E. C. Lee, D. Kim, P. Jurecka, P. Tarakeshwar, P. Hobza, and K. S. Kim, The Journal of Physical Chemistry A **111**, 3446 (2007).
[129] J. F. Pereira, L. A. Flores, H. Wang, and R. D. Rogers, Chemistry–A European Journal **20**, 15482 (2014).
[130] T. Shimomura, T. Takamuku, and T. Yamaguchi, The Journal of Physical Chemistry B **115**, 8518 (2011).
[131] H. Shirota, S. Kakinuma, Y. Itoyama, T. Umecky, and T. Takamuku, The Journal of Physical Chemistry B **120**, 513 (2016).
[132] S. Li, G. Saielli, and Y. Wang, Physical Chemistry Chemical Physics **20**, 22730 (2018).
[133] S. Li, N. Safari, G. Saielli, and Y. Wang, J Phys Chem B **124**, 7929 (2020).
[134] S. S. Bystrov, V. V. Matveev, A. V. Egorov, Y. S. Chernyshev, V. A. Konovalov, V. Balevičius, and V. I. Chizhik, The Journal of Physical Chemistry B **123**, 9187 (2019).
[135] J. Carrete, T. Mendez-Morales, O. Cabeza, R. M. Lynden-Bell, L. J. Gallego, and L. M. Varela, The Journal of Physical Chemistry B **116**, 5941 (2012).
[136] P. D'Angelo, A. Zitolo, G. Aquilanti, and V. Migliorati, The Journal of Physical Chemistry B **117**, 12516 (2013).
[137] M. H. Ghatee and A. R. Zolghadr, The Journal of Physical Chemistry C **117**, 2066 (2013).
[138] G. A. Hegde, V. S. Bharadwaj, C. L. Kinsinger, T. C. Schutt, N. R. Pisierra, and C. M. Maupin, The Journal of Chemical Physics **145**, 064504 (2016).
[139] T. Koishi, The Journal of Physical Chemistry B **122**, 12342 (2018).
[140] V. Migliorati, A. Zitolo, and P. D'Angelo, The Journal of Physical Chemistry B **117**, 12505 (2013).

[141] M. Moreno, F. Castiglione, A. Mele, C. Pasqui, and G. Raos, The Journal of Physical Chemistry B **112**, 7826 (2008).
[142] S. D. Nickerson, E. M. Nofen, H. Chen, M. Ngan, B. Shindel, H. Yu, and L. L. Dai, The Journal of Physical Chemistry B **119**, 8764 (2015).
[143] C. Schröder, T. Rudas, G. Neumayr, S. Benkner, and O. Steinhauser, The Journal of chemical physics **127**, 234503 (2007).
[144] J. M. Vicent‐Luna, D. Dubbeldam, P. Gómez‐Álvarez, and S. Calero, ChemPhysChem **17**, 380 (2016).
[145] W. Jiang, Y. Wang, and G. A. Voth, The Journal of Physical Chemistry B **111**, 4812 (2007).
[146] S. Ohta, A. Shimizu, H. Abe, N. Hatano, Y. Ima, and Y. Yoshimura, Open Journal of Physical Chemistry **1**, 70 (2011).
[147] H. Abe, T. Takekiyo, M. Shigemi, Y. Yoshimura, S. Tsuge, T. Hanasaki, K. Ohishi, S. Takata, and J.-i. Suzuki, The Journal of Physical Chemistry Letters **5**, 1175 (2014).
[148] H. Abe, T. Takekiyo, Y. Yoshimura, K. Saihara, and A. Shimizu, ChemPhysChem **17**, 1136 (2016).
[149] K. Saihara, Y. Yoshimura, S. Ohta, and A. Shimizu, Scientific reports **5**, 1 (2015).
[150] C. E. Bernardes, M. E. Minas da Piedade, and J. N. Canongia Lopes, The Journal of Physical Chemistry B **115**, 2067 (2011).
[151] J. N. Canongia Lopes and A. A. Pádua, The Journal of Physical Chemistry B **110**, 3330 (2006).
[152] R. Hayes, G. G. Warr, and R. Atkin, Chemical reviews **115**, 6357 (2015).
[153] K. Shimizu, C. E. Bernardes, and J. N. Canongia Lopes, The Journal of Physical Chemistry B **118**, 567 (2014).
[154] K. Shimizu, M. F. C. Gomes, A. A. Pádua, L. P. Rebelo, and J. N. C. Lopes, Journal of Molecular Structure: THEOCHEM **946**, 70 (2010).
[155] O. Russina, F. Lo Celso, N. V. Plechkova, and A. Triolo, The Journal of Physical Chemistry Letters **8**, 1197 (2017).
[156] T. Cosby, U. Kapoor, J. K. Shah, and J. Sangoro, J Phys Chem Lett **10**, 6274 (2019).
[157] S. Fuladi, H. Gholivand, A. Ahmadiparidari, L. A. Curtiss, A. Salehi-Khojin, and F. Khalili-Araghi, J Phys Chem B **125**, 7024 (2021).